\documentclass[10pt,twoside,leqno,twocolumn]{article}
\usepackage{ltexpprt}
\usepackage[letterpaper,margin=1in]{geometry}

\usepackage{amsfonts,amsmath}
\usepackage{cite}
\usepackage{textcomp}
\usepackage{xcolor}
\usepackage{algorithm}
\usepackage[noend]{algpseudocode}
\usepackage{booktabs}
\usepackage{csquotes}
\usepackage{enumerate}
\usepackage{multirow}
\usepackage[figuresleft]{rotating}
\usepackage[autolanguage]{numprint}
\usepackage{graphicx}
\usepackage[utf8]{inputenc}
\usepackage[locale=US]{siunitx}
\usepackage{subcaption}
\usepackage{textcomp}
\usepackage[square,sort,comma,numbers]{natbib}
\usepackage{url}
\usepackage{tikz}
\usepackage{xspace}
\usepackage{hyperref}
\usepackage{cleveref}

\definecolor{light-gray}{gray}{0.93}
\usepackage[backgroundcolor=light-gray]{todonotes}

\newcommand{\distrSpeedupCplx}{\numprint{12.2}\times\xspace}
\newcommand{\distrSpeedupRoad}{\numprint{11.5}\times\xspace}
\newcommand{\parSpeedupCplx}{\numprint{8.7}\times\xspace}
\newcommand{\parSpeedupRoad}{\numprint{9.2}\times\xspace}
\newcommand{\ustMaxAbsFour}{\numprint{0.14}\xspace}

\newcommand{\jltJuliaMaxAbsOne}{\numprint{0.15}\xspace}
\newcommand{\ustJuliaFourOneSpeedup}{\numprint{397.5}\times\xspace}
\newcommand{\ustJLTSpeedupCplx}{\numprint{7.6}\times\xspace}
\newcommand{\ustJLTSpeedupRoad}{\numprint{1.9}\times\xspace}

\newcommand{\ie}{i.\,e.,\xspace}
\newcommand{\eg}{e.\,g.,\xspace}
\newcommand{\wrt}{w.\,r.\,t.\xspace}

\newcommand{\etal}{et al.\xspace}

\newcommand{\E}{{\ensuremath{\mathbb{E}}}}
\newcommand{\Oh}{\ensuremath{\mathcal{O}}}
\newcommand{\argmin}{\operatorname{argmin}}
\newcommand{\argmax}{\operatorname{argmax}}

\newcommand{\diag}[1]{\operatorname{diag}(#1)}
\newcommand{\diam}[1]{\operatorname{diam}(#1)}

\newcommand{\mat}[1]{\mathbf{#1}}
\newcommand{\myvec}[1]{\mathbf{#1}}
\newcommand{\ment}[3]{\mathbf{#1}[#2,#3]}
\newcommand{\vent}[2]{\myvec{#1}(#2)}
\newcommand{\Lpinv}{\mat{L_{G}}^\dagger}
\newcommand{\Pomega}{\mat{\Omega}}
\newcommand{\Pomegaalpha}{\mat{\Omega}_{\alpha}}

\newcommand{\Aug}{G_\star}
\newcommand{\Lap}{\mat{L}}
\newcommand{\AugLap}{\mat{L}_\star}
\newcommand{\msub}[2]{\left(#1\right)_{-#2}}
\newcommand{\effres}[2]{\myvec{r}(#1,#2)}
\newcommand{\effresG}[3]{\myvec{r}_{#1}(#2,#3)}
\newcommand{\fdistpalpha}[2]{\myvec{\rho}_\alpha(#1,#2)}
\newcommand{\fdistp}[2]{\myvec{\rho}(#1,#2)}
\newcommand{\fdistu}[1]{\myvec{\rho}(#1)}

\newcommand{\fdistinv}[1]{\myvec{f_{\alpha}}(#1)}
\newcommand{\trace}[1]{\operatorname{tr}(#1)}
\newcommand{\uvec}[1]{\mathbf{e}_{#1}}
\newcommand{\onesvec}{\mathbf{1}}
\newcommand{\ecc}{\operatorname{ecc}}

\newcommand{\poly}[1]{\operatorname{poly}(#1)}

\renewcommand{\epsilon}{\varepsilon}
\newcommand{\pivot}{u^\star}

\newcommand{\jltCpp}{\tool{JLT-CPP}\xspace}
\newcommand{\jltKyng}{\tool{JLT-Julia}\xspace}
\newcommand{\algo}{\tool{UST}\xspace}

\newcommand{\lamgTol}{10^{-9}\xspace}

\newcommand{\tool}[1]{\textsf{#1}}

\newtheorem{definition}{Definition}

\newif{\ifblind}
\blindfalse
\newif{\ifcomments}
\commentstrue

\newif{\ifarxiv}
\arxivtrue

\ifarxiv
\newcommand{\arxivOrCamera}[2]{#1}
\else
\newcommand{\arxivOrCamera}[2]{#2}
\fi

\ifcomments
\newcommand{\changed}[1]{#1}
\newcommand{\hmey}[1]{{\color{red}[HM: #1]}}
\newcommand{\ea}[1]{{\color{orange}[EA: #1]}}
\newcommand{\avdg}[1]{{\color{olive}[AvdG: #1]}}
\newcommand{\mpre}[1]{{\color{brown}[MP: #1]}}
\else
\newcommand{\changed}[1]{#1}
\newcommand{\hmey}[1]{}
\newcommand{\ea}[1]{}
\newcommand{\avdg}[1]{}
\newcommand{\mpre}[1]{}
\fi

\ifblind
\newcommand{\ours}{the algorithm proposed by Angriman \etal} 

\else
\newcommand{\ours}{our algorithm for electrical closeness}

\fi
\blindtrue

\newcommand*\samethanks[1][\value{footnote}]{\footnotemark[#1]}

\begin{document}

\author{Alexander van der Grinten\thanks{Dept.\ of Computer Science, Humboldt-Universit\"at zu Berlin,
Unter den Linden 6, D-10099 Berlin.
\texttt{\{avdgrinten, angrimae, predarim, meyerhenke\}@hu-berlin.de}}
\and Eugenio Angriman\samethanks[2]
\and Maria Predari\samethanks[2]
\and Henning Meyerhenke\samethanks[2]}

\newcommand{\ourtitle}{New Approximation Algorithms for Forest Closeness Centrality -- 
for Individual Vertices and Vertex Groups}

\title{\ourtitle \thanks{
This work is partially supported by German Research Foundation (DFG) grant ME 3619/3-2
within Priority Programme 1736 and by DFG grant ME 3619/4-1.
}
}
\date{}

\maketitle
\arxivOrCamera{}{\fancyfoot[R]{\scriptsize{Copyright \textcopyright\ 2021 by SIAM\\
Unauthorized reproduction of this article is prohibited}}}

\begin{abstract}
The emergence of massive graph data sets requires fast mining algorithms.
Centrality measures to identify important vertices belong to the most popular analysis methods
in graph mining. A measure that is gaining attention is forest closeness centrality; 
it is closely related to electrical measures using current flow
but can also handle disconnected graphs.
Recently, [Jin et al., ICDM'19] proposed an algorithm
to approximate this measure probabilistically. 
Their algorithm processes small inputs quickly, but does not scale well 
beyond hundreds of thousands of vertices.

In this paper, we first propose a different approximation
algorithm; it is up to two orders of magnitude
faster and more accurate in practice.
Our method exploits the strong connection between uniform spanning 
trees and forest distances by adapting and extending recent approximation algorithms
for related single-vertex problems.
This results in a nearly-linear time algorithm
with an absolute probabilistic error guarantee.
In addition, we are the first to consider the problem of finding
an optimal \emph{group} of vertices \wrt\ forest closeness. 
We prove that this latter problem is NP-hard;
to approximate it, we adapt a greedy algorithm by [Li \etal, WWW'19],
which is based on (partial) matrix inversion.
Moreover, our experiments show that on disconnected graphs,
group forest closeness
outperforms existing centrality measures in the context
of semi-supervised vertex classification.

\end{abstract}

\section{Introduction}
\label{sec:intro}
%
Massive graph data sets with millions of edges (or more) have become abundant.
Today, applications come from many different scientific and
commercial fields~\cite{newman2018networks,barabasi2016network}.
Network analysis algorithms shall uncover non-trivial relationships
between vertices or groups of vertices in these data.
One popular concept used in network analysis is \emph{centrality}.
Centrality measures assign to each vertex (or edge) a score
based on its structural importance; this allows to rank the vertices and to identify the important \changed{ones~\cite{DBLP:journals/im/BoldiV14,DBLP:conf/kdd/WhiteS03}.}
\changed{Measures that capture not only local graph properties are often more meaningful, yet relatively
expensive to compute~\cite{Grinten2020ScalingUN}. 
Also,} different applications may re\-quire different centrality measures, none is universal.

\changed{Algebraic measures such as random-walk betweenness, electrical closeness
(see Refs.\ in~\cite{angrimanPGM20,Grinten2020ScalingUN}), and} 
\emph{forest closeness centrality}~\cite{Zhang19} are gaining \changed{increasing} attention.
\changed{Forest closeness} is based on forest distance, which was
introduced by Chebotarev and Shamis~\cite{Chebotarev00} to account not only for shortest 
paths.\footnote{Instead, all paths are taken into account, but shorter ones are more important.
This notion of distance/proximity has many applications in graph/data mining and beyond~\cite{Chebotarev00}.}
Moreover, it applies to disconnected graphs as well.
\changed{ In sociology, forest distances are shown to better capture more than one sensitive relationship index,
  such as social proximity and group cohesion~\cite{chebotarev06matrixforest}.}
Consequently, forest closeness centrality has two main advantages \changed{over many other
centrality} measures~\cite{Zhang19}: (i) by taking not only shortest paths into account,
it has a high discriminative power and (ii) unlike \changed{related algebraic measures such as the above},
it can handle disconnected graphs out of the box.

Recently, Jin \etal~\cite{Zhang19} provided an approximation algorithm for forest closeness centrality
with nearly-linear time complexity. Their algorithm uses the Johnson-Lindenstrauss transform (JLT) 
and fast linear solvers; it 
can handle much larger inputs than what was doable before,
but is still time-consuming. For example, graphs with $\approx$1M vertices and $\approx$2-3M edges
require more than $2.3$ or $4.7$ \emph{hours} for a reasonably accurate ranking in their study~\cite{Zhang19}. 
Obviously, this hardly scales to massive graphs with $> 50$M edges; corresponding applications 
would benefit significantly from faster approximation methods.

To this end, we devise new approximation algorithms for two problems:
first, for the individual forest closeness centrality value of each node --
by adapting uniform spanning tree techniques from recent related work on electrical closeness centrality~\cite{angrimanPGM20,barthelm2019estimating}.
In a next step, we consider \emph{group} forest closeness centrality, where one
seeks a set of vertices that is central jointly. To the best of our knowledge,
we are the first to address the group case for this centrality measure. We prove that group
forest closeness is $\mathcal{NP}$-hard and adapt the greedy algorithm by Li \etal~\cite{DBLP:conf/www/0002PSYZ19}
to this problem.
Our experiments on common benchmark graphs show that our algorithm for ranking individual
vertices is always substantially faster than Jin \etal's~\cite{Zhang19} -- for sufficiently large
networks by one (better accuracy) to two (similar accuracy) orders of magnitude in a sequential setting.
Our new algorithm can now rank all vertices in networks of up to $334$M edges with reasonable accuracy
in less than 20 minutes if executed in an MPI-parallel setting.
Also, experiments on semi-supervised vertex classification
demonstrate that our new group forest closeness measure
improves upon existing measures in the case of disconnected graphs.

\section{Definitions and Notation}
\label{sec:prelim}
As input we consider finite and simple undirected graphs $G = (V, E, \myvec{w})$ with $n$ vertices,
$m$ edges, and edge weights $\myvec{w} \in \mathbb{R}_{\geq 0}^{m}$.
By $\Lap$ we denote the Laplacian matrix of $G$, defined as $\Lap
= \mat{diag}(\deg_G(1), \ldots, \deg_G(n)) - \mat{A}_G$, where $\mat{A}_G$ denotes
the (weighted) adjacency matrix of $G$ and $\deg_G(v)$ the (weighted) degree of vertex $v$.

\paragraph{Closeness centrality.}
Let $d(u, v)$ denote the graph distance in $G$.
The \emph{farness} of a vertex $u$ is defined as
$f^d(u) := \sum_{v \neq u} d(u, v)$,
\ie up to a scaling factor of $\frac 1 n$, the farness of $u$ quantifies the average
distance of $u$ to all other vertices.
Given this definition, the \emph{closeness centrality}
of $u$ is defined as $C^d(u) := \frac n{f^d(u)}$.
Closeness is a widely used centrality measure;
the higher the numerical value of $C^d(u)$ is, the more central is $u$ within the graph.
It is often criticized for mapping the vertex scores into a rather narrow interval~\cite{newman2018networks}.

\paragraph{Forest Distance / Closeness.}
Forest distance generalizes the common graph distance and takes not only shortest paths
into account~\cite{Chebotarev00}. It is expressed in terms of the (parametric) forest matrix of a graph $G$ defined as
$\Pomega := \Pomegaalpha := (\alpha\Lap + \mat{I})^{-1}$,
where $\mat{I}$ is the identity matrix and $\alpha > 0$ controls the importance
of short vs long paths between vertices
(some papers prefer the expression $(\Lap + \alpha \mat{I})^{-1}$,
which is equivalent to $\Pomegaalpha$ up to scaling;
non-parametric variants of forest closeness fix
$\alpha$ to $1$~\cite{chebotarev2006proximity}):
\begin{definition}[Forest distance~\cite{Chebotarev00}]
  \label{def:forest-dist-pair}
  The forest distance $\fdistp{u}{v}$ for a vertex pair $(u,v)$ is defined as:
  \begin{equation}
      \label{eq:forest-dist-pair}
    \begin{split}
    \fdistp{u}{v} & := \fdistpalpha{u}{v} := (\uvec{u} - \uvec{v})^T \Pomegaalpha (\uvec{u} - \uvec{v})
    \\ & = \ment{\Pomegaalpha}{u}{u} + \ment{\Pomegaalpha}{v}{v} -2\ment{\Pomegaalpha}{u}{v}.
    \end{split}
  \end{equation}
\end{definition}
\changed{Chebotarev and Shamis~\cite{Chebotarev00} show} that forest distance is a metric and list other desirable properties.
The name \emph{forest} distance stems from the fact that an entry $\ment{\Pomega}{u}{v}$ equals the fraction 
of spanning rooted forests in $G$ in which $u$ and $v$ belong to the same tree, see~\cite{Zhang19}.
Forest distance closeness centrality, or forest closeness for short, then uses forest distances
instead of the usual graph distance in the sum over all other vertices:
\begin{definition}[Forest closeness~\cite{Chebotarev00}]
  \label{forest-dist-vertex}
  The \emph{forest farness} $\fdistu{u}$ of a vertex $u$
   is defined as $\fdistu{u} :=  \sum_{v \in V \setminus \{u\}}\fdistp{u}{v}$.
	Likewise, the \emph{forest distance closeness centrality} of $u$ is defined as:
     $\fdistinv{u} := \frac{n}{\fdistu{u}}$.
\end{definition}

To simplify notation and when clear from the context, we often omit $\alpha$ in the following.

\paragraph{Effective Resistance and Electrical Closeness.}
As already realized by Chebotarev and Shamis~\cite{Chebotarev00}, there is a close
connection between forest distance and effective resistance, a.\,k.\,a. resistance distance
(more details on this connection in Section~\ref{sub:connection-forest-resistance}).
Effective resistance is a pairwise metric on the vertex set of a graph
and also plays a central role in several centrality 
measures~\cite{teixeira2013spanning, DBLP:conf/stacs/BrandesF05}.
The notion of effective resistance comes from viewing $G$
as an electrical circuit in which each edge $e$ is a resistor
with resistance $1/\vent{w}{e}$.
Following fundamental electrical laws, the effective resistance $\effres{u}{v}$
between two vertices $u$ and $v$ (that may or may not share an edge)
is the potential difference between $u$ and $v$ when a unit
of current is injected into $G$ at $u$ and extracted at $v$.
\changed{Effective resistance is also proportional to hitting times of 
random walks~\cite{DBLP:books/daglib/0009415} and thus has connections to Markov chains.}
Computing the effective resistance $\effres{u}{v}$
of a vertex pair $(u,v) \in V \times V $ can be done by means of the Laplacian
pseudoinverse $\Lpinv$ as
\changed{
\begin{equation}
  \label{eq:eff-res}
  \effres{u}{v} = \ment{\Lpinv}{u}{u} + \ment{\Lpinv}{v}{v} - 2 \ment{\Lpinv}{u}{v}
\end{equation}
}(or by solving a Laplacian linear system).
Given the definition of $\effres{u}{v}$, one obtains the well-known
definition of \emph{electrical closeness} by replacing
$\fdistu{u}$ by $\effres{u}{v}$
in Definition~\ref{forest-dist-vertex}.
Electrical closeness (aka \emph{current-flow closeness} or \emph{information centrality})
has been widely studied (see \eg~\cite{DBLP:conf/stacs/BrandesF05,DBLP:conf/www/0002PSYZ19,DBLP:conf/siamcsc/BergaminiWLM16,Grinten2020ScalingUN}),
but only in the context of connected graphs.

\section{Related Work}
\label{sc:rel-work}
The most relevant algorithmic work regarding forest closeness
was proposed by Jin \etal~\cite{Zhang19}, who presented an $\epsilon$-approximation algorithm for
forest distance and forest closeness for all graph nodes.
The authors exploit the Johnson-Lindenstrauss lemma~\cite{johnson1984extensions}, thus use random projections
and rely on fast Laplacian solvers~\cite{CohenKyng14}
to avoid matrix inversions. 
The algorithm has a running time of $\Oh(m\epsilon^{-2}\log^{2.5}{n}\log(1/\epsilon)\poly{\log\log n})$
and provides a $(1\pm \epsilon)$-approximation guarantee with high probability (assuming an exact Laplacian solver).
In practice, as mentioned above, their approach takes $> 2$ hours
on graphs with $\approx$1M vertices and $\approx$2-3M edges for a reasonably accurate ranking. 
\changed{Our aim is a better algorithmic solution for forest centrality by
leveraging our recent results on the approximation of
the diagonal entries of \changed{$\Lpinv$}~\cite{angrimanPGM20}.
The latter exploits the connection to effective resistances
and electrical closeness and is stated here for completeness:}
\begin{proposition}[\cite{angrimanPGM20}]
\label{effres:time-complexity}
Let $G = (V,E)$ be an undirected and weighted graph with diameter $\diam{G}$
and volume $\operatorname{vol}(G)$.
There is an algorithm that computes with probability $1-\delta$ an approximation of $\diag{\Lpinv}$
with absolute error $\pm \epsilon$ in expected time
$\Oh(\operatorname{vol}(G) \cdot \ecc^3(u) \cdot \epsilon^{-2} \cdot \log(\operatorname{vol}(G)/\delta))$
\changed{, where $\ecc(u)$ is the eccentricity of a selected node $u$}.
\end{proposition}
That algorithm exploits three major insights:
(i) to compute the electrical closeness of a node $u$, one only needs $\ment{\Lpinv}{u}{u}$
and the trace of $\Lpinv$;
(ii) after obtaining the $u$-th column of $\Lpinv$ (by solving one Laplacian linear system)
and all effective resistances $\effres{u}{v}$ between $u$ and all $v$,
the remaining elements of $\operatorname{diag}(\Lpinv)$ can be calculated via Eq.~(\ref{eq:eff-res}),
(iii) effective resistances can be approximated by sampling uniform spanning trees (USTs), \eg with Wilson's algorithm~\cite{Wilson:1996:GRS:237814.237880},
by exploiting Kirchhoff's theorem.
For our purposes, we can state it as the effective resistance of an edge $\{u,v\} \in E$ being the probability that 
$\{u,v\}$ is in a spanning tree drawn uniformly at random from all spanning trees of $G$  (comp.~\cite{DBLP:books/daglib/0009415}).

The algorithm proposed in this paper for approximating individual centrality scores is 
based on the above insights, transfers them to a different
graph and provides a new analysis with an improved running time for the case at hand.

Barthelm\'e \etal~\cite{barthelm2019estimating} proposed an algorithm that uses techniques similar to
the ones in Ref.~\cite{angrimanPGM20} to estimate inverse traces
that arise in regularized optimization problems.
Their algorithm is based on uniform spanning forests,
also sampled with Wilson's algorithm.
Finally, for the group centrality case, the most relevant algorithm is Li \etal's~\cite{DBLP:conf/www/0002PSYZ19};
it employs JLT and fast Laplacian solvers to approximate group electrical closeness
centrality in nearly-linear time.

\section{Forest Closeness of Individual Vertices}
\label{sec:algorithm}
By definition, forest closeness for a vertex $u$ can be computed from
all forest distances $\fdistp{u}{v}$, $v \in V \setminus \{u\}$,
\eg by matrix inversion. Yet, inversion takes cubic time in practice and is thus impractical for large graphs.
Hence, we exploit a relation between forest distance and effective resistance
to approximate the forest farness more efficiently than existing approximation algorithms.
By adapting \ours~\cite{angrimanPGM20}, we obtain 
an algorithm with a (probabilistic) \changed{additive} approximation guarantee of $\pm \epsilon$;
it runs in nearly-linear (in $m$) expected time.

\subsection{From Forest Farness to Electrical Farness (And Back Again).}
\label{sub:connection-forest-resistance}
As mentioned, we exploit a result that relates
forest distances to effective resistances.
This requires the creation of an \emph{augmented} graph $\Aug := G_{\star, \alpha} := (V',E')$
from the original graph $G = (V,E)$.
To this end, a new \emph{universal vertex} $\pivot$ is added to $G$,
such that $V' =  V \cup \{\pivot\}$ and  $E' = E \cup \{\pivot, v\}, ~\forall v \in V$.
In particular, $\pivot$ is connected
to all other vertices of $\Aug$ with edges of
weight one.
Furthermore, the weights of all edges in $E'$ that belong to $E$ are
\changed{multiplied} by $\alpha$. 

\begin{proposition}[comp.\ Ref.~\cite{Chebotarev00}]
  \label{forest-resistance}
For a weigh\-ted graph $G = (V,E)$ and any vertex pair $ (v_1,v_2) \in V \times V$,
the forest distance $\fdistp{v_1}{v_2}$ in $G$ equals
the effective resistance $\effres{v_1}{v_2}$ \changed{in the augmented graph $\Aug$.}
\end{proposition}

The full proof of Proposition~\ref{forest-resistance} can be found
in Ref.~\cite{Chebotarev00}. Nevertheless, we provide here an explanation of
why the above proposition holds.
Recall that the effective resistance between any two vertices
of $G$ is computed by means of $\Lpinv$,
while the forest distances of the same pair are computed by means of
the forest matrix of $G$, $\mat{\Pomega} = (\alpha\Lap+\mat{I})^{-1}$.
When calculating the effective resistance in $\Aug$, we use its Laplacian matrix
$\AugLap$, which consists of a block matrix corresponding to $(\alpha\Lap + \mat{I})$ and
an additional row and column that corresponds to the universal
vertex $\pivot$.
It turns out that the Moore-Penrose pseudoinverse of $\AugLap$ is the block matrix
that consists of
$\Pomega$ with an additional row and column corresponding to $\pivot$~\cite{Chebotarev00}.
Thus, \changed{$\ment{\Pomega}{\pivot}{\pivot} + \ment{\Pomega}{v}{v} - 2\ment{\Pomega}{\pivot}{v} $ equals
$\ment{\AugLap^\dagger}{\pivot}{\pivot} + \ment{\AugLap^\dagger}{v}{v} - 2\ment{\AugLap^\dagger}{\pivot}{v} $},
which corresponds to the pairwise effective resistance \changed{$\effres{\pivot}{v}$} in $\Aug$.

\begin{corollary}
\label{fcl-elcl}
Forest closeness in graph $G$ equals electrical closeness in the augmented graph $\Aug$.
\end{corollary}


\subsection{Forest Farness Approximation Algorithm.}
\label{sub:new-forest-algo}
As mentioned, our new algorithm for forest closeness exploits previous algorithmic results for 
approximating $\operatorname{diag}(\Lpinv)$ and electrical
closeness. To do so, we rewrite forest farness $\fdistu{v}$ following Ref.~\cite{Merris98}:
\begin{small}
\begin{align}
\label{eq-fdistu-tr}
\begin{split}
  \fdistu{v} & =  n \cdot \ment{\Pomega}{v}{v} + \trace{\Pomega} - 2 \sum_{w \in V} \ment{\Pomega}{v}{w} \\
  & = n \cdot \ment{\Pomega}{v}{v} + \trace{\Pomega} - 2,
\end{split}
\end{align}
\end{small}
where the last equation holds since $\Pomega$ is doubly stochastic
($\ment{\Pomega}{v}{v} = 1 - \sum_{w \neq v}\ment{\Pomega}{v}{w}$)~\cite{Merris98}.
From Eq.~(\ref{eq-fdistu-tr}) it is clear that we only require the diagonal elements
of $\Pomega$ to compute $\fdistu{v}$ for any $v \in V$.
We approximate the diagonal elements of $\Pomega$ with Algorithm~\ref{alg:approx-diag-omega},
whose main idea is to sample uniform spanning trees (USTs) to approximate $\diag{\AugLap^{\dagger}}$:
\begin{enumerate}
\item We build the augmented graph $\Aug$ (Line~\ref{line:build-augmented-graph}) and let the universal vertex
  $\pivot$ of $\Aug = (V',E')$ be the so-called \emph{pivot vertex}
  (Line~\ref{line:set-pivot}) -- due to its optimal eccentricity of $1$. 
  Later, we compute the column of $\Pomega$ that corresponds
  to $\pivot$, $\ment{\Pomega}{:}{\pivot}$, by solving the Laplacian linear system
  $\AugLap \myvec{x} = \uvec{\pivot} - \frac{1}{n+1}\cdot \onesvec$ (Line~\ref{line:solve-system}).
  The solver's accuracy is controlled by $\eta$, which is set in Line~\ref{line:set-eta}
  ($\kappa$ is used to trade the accuracy of the solver with the accuracy of the following sampling step).
\item We sample $\tau$ USTs in $\Aug$ with Wilson's algorithm~\cite{Wilson:1996:GRS:237814.237880}
  (also see
  \changed{\arxivOrCamera{\Cref{alg:sampling-ust} in \Cref{sec:app-algorithmic-details}}{Algorithm 3 in the full version~\cite{full-version}}}),
  where the sample size $\tau$ is yet to be determined.
  With this sample we approximate the effective resistance $\effresG{\Aug}{\pivot}{v}$ for all $v \in V$
  (Lines~\ref{line:ust-sampling-loop}-\ref{line:ust-sampling}). More precisely, if an edge $\{\pivot,v\}$ appears
  in the sampled tree, we increase $R[v]$ by $1$ (unweighted case) or by the weight of the current 
  tree (weighted case) -- and later ``return'' $R[v] / \tau$ (unweighted case) or the relative total weight of 
  all sampled trees (weighted case) that contain edge $\{\pivot,v\}$ in Line~\ref{line:diag-remain}.
\item We compute the remaining $\ment{\Pomega}{v}{v}$ for $v \in V $ in Lines~\ref{line:fill-loop} and~\ref{line:diag-remain}
  following Eqs.~(\ref{eq:forest-dist-pair}) and~(\ref{eq:eff-res}):
\begin{align*}
	\label{eq:diag-comp}
	\ment{\Pomega}{v}{v} & = \fdistp{\pivot}{v} - \ment{\Pomega}{\pivot}{\pivot} + 2 \ment{\Pomega}{v}{\pivot} \\
	& = \effresG{\Aug}{\pivot}{v}- \ment{\Pomega}{\pivot}{\pivot} + 2 \ment{\Pomega}{v}{\pivot},
\end{align*}
where $\effresG{\Aug}{\pivot}{v}$ is then approximated by $R[v] / \tau$
(the weighted case is handled as described above).
\end{enumerate}


\begin{algorithm}[bt]
  \begin{algorithmic}[1]
    \begin{small}
      \Function{ApproxDiagForestMatrix}{$G$, $\alpha$, $\epsilon$, $\delta$}
      \State \textbf{Input:} Undir.\ graph $G = (V, E)$, control parameter $\alpha$,
      error bound $0 < \epsilon < 1$, probability $0 < \delta < 1$
      \State \textbf{Output:} $\diag{\widetilde{\Pomega}}$, \ie an $(\epsilon, \delta)$-approximation of $\diag{\Pomega}$
      \State Create augmented graph $\Aug = (V',E')$ as described in Proposition~\ref{forest-resistance}; compute $\operatorname{vol(G)}$ and $c$ \label{line:build-augmented-graph}
      \Comment{$\Oh(m+n)$}
      \State $\pivot \gets $ universal vertex of $\Aug$  \label{line:set-pivot}
      \State Pick constant $\kappa \in (0, 1)$ arbitrarily  \label{line:pick-kappa}
      \State $\eta \gets \frac{\kappa \epsilon}{6 \sqrt{\alpha (c+2) \operatorname{vol}(G)}}$ \label{line:set-eta}
      \State $\tau \gets \lceil \log(2m/\delta) / 2(1-\kappa)^2\epsilon^2\rceil$

    \For{$i \gets 1$ to $\tau$}\label{line:ust-sampling-loop}
      \Comment {\small $\tau$ times}
      \State $R \gets$ \textsc{SamplingUST}($\Aug$, $u$)
      \label{line:wilson}
      \Comment{\small $\Oh(\alpha \operatorname{vol}(G) + n)$}\label{line:ust-sampling}
    \EndFor 
    
    \State Solve \changed{ $\AugLap \myvec{x} = \uvec{\pivot} - \frac{1}{n+1}\cdot \onesvec$ }for $\myvec{x}$
    \Comment {\small accuracy: $\eta$, $\tilde{\Oh}(m \log^{1/2} n \log(1/\eta))$} \label{line:linear-system} \label{line:solve-system}
    \For{$v \in V$} \label{line:fill-loop}
    \Comment {\small All iterations: $\Oh(n)$}
      \State \changed{ $\ment{\widetilde{\Pomega}}{v}{v} \gets R[v] / \tau - \vent{x}{\pivot} + 2 \vent{x}{v}$} \label{line:diag-remain}
      \Comment{unweighted case, for weighted see text}
      \EndFor \label{line:end-fill}
      \State \textbf{return} $\operatorname{diag}(\widetilde{\Pomega})$
    \EndFunction
    \end{small}
  \end{algorithmic}
  \caption{Approximation algorithm for $\diag{\Pomega}$}
  \label{alg:approx-diag-omega}
\end{algorithm}


By using $\Aug$ and thus a universal vertex $\pivot$ as pivot, there are several noteworthy changes
compared to the algorithm in Ref.~\cite{angrimanPGM20}. First, the graph $\Aug$ has constant
diameter and the vertex $\pivot$ constant eccentricity $1$. This will be important for our refined
running time analysis. Second, the approximation of the effective resistances can be simplified:
while Ref.~\cite{angrimanPGM20} requires an aggregation along shortest paths, we notice that
here $\pivot$ and all other vertices are connected by paths of one edge only; thus,
the relative frequency of an edge $\{\pivot,v\}$ in the UST sample for $\Aug$ is sufficient here
for our approximation:

\begin{proposition}
\label{prop:unbiased}
Let $\pivot$ be the universal vertex in $\Aug$.
Then, for any edge $\{\pivot,v\} \in E'$ holds: its relative frequency (or weight) in the UST sample
is an unbiased estimator for $\effresG{\Aug}{\pivot}{v}$.
\end{proposition}
The proof of Proposition~\ref{prop:unbiased} relies
on Kirchhoff's theorem (see~\cite[Ch.~II]{DBLP:books/daglib/0009415})
and can be found in \changed{\arxivOrCamera{\Cref{apx:technical-proofs}}{the full version
of this paper~\cite{full-version}}}.

As we will see in our main algorithmic result (Theorem~\ref{thm:time-approx}),
Algorithm~\ref{alg:approx-diag-omega} is not only an unbiased estimator, 
but even provides a probabilistic approximation guarantee. To bound its running
time, we analyze Wilson's algorithm for generating a UST first.

\begin{proposition}
\label{prop:ust-wilson-time}
  For an undirected graph $G$ with constant diameter, each call to Wilson's algorithm
  on $\Aug$ (in 
  Line~\ref{line:ust-sampling})
  takes $\Oh(\alpha \operatorname{vol}(G) + n)$ expected time,
  where $\operatorname{vol}(G) = \sum_{v \in V} \deg(v)$ is the (possibly weighted) volume of $G$.
\end{proposition}
The proof of Proposition~\ref{prop:ust-wilson-time} can be found
in \changed{\arxivOrCamera{\Cref{apx:technical-proofs}}{the full version~\cite{full-version}}}.
Note that in the case of unweighted graphs with $\alpha = 1$ and $m = \Omega(n)$ (which is not uncommon in
our context, see for example Ref.~\cite{Zhang19}), we obtain
a time complexity of $\Oh(m)$ (the volume is $2m$ by the handshake lemma).
Taking all the above into account, we arrive at our main algorithmic result on running time
and approximation bounds of Algorithm~\ref{alg:approx-diag-omega}. 
The result and its proof are adaptations of Theorem~3 in Ref.~\cite{angrimanPGM20}.
When considering forest (as opposed to electrical) closeness centrality, we exploit the constant diameter of $\Aug$
and improve the time by a factor of $(\ecc(u))^3$, \changed{ where $u$ is a selected pivot node}.
This expression is $\Oh(\log^3 n)$ for the small-world graphs in the focus of Ref.~\cite{angrimanPGM20}
(but can be larger for general graphs).  In the following,
$\tilde{\Oh}(\cdot)$ hides polyloglog factors from the linear solver~\cite{CohenKyng14}.
\begin{theorem}
\label{thm:time-approx}
Let $\frac{n}{\alpha \cdot \operatorname{vol}(G)}$ be bounded from above by a constant\footnote{The condition ensures that the algorithm is not affected by unduly heavy additional edges
to $\pivot$. If the condition is met, the graph edges still play a reasonable role in the distances
and in the UST computations.} and
let $0 < \epsilon, \delta < 1$. 
Then, with probability $1-\delta$, Algorithm~\ref{alg:approx-diag-omega}
computes an approximation of $\diag{\Pomega}$ with absolute error $\pm \epsilon$ in (expected) time 
$\tilde{\Oh}((m \log^{1/2} n \log(\sqrt{\alpha \operatorname{vol}(G)} / \epsilon))) + 
\Oh(\log(n / \delta) / \epsilon^2 \cdot \alpha \operatorname{vol}(G))$.
\end{theorem}
Theorem~\ref{thm:time-approx} is proved in \changed{\arxivOrCamera{\Cref{apx:technical-proofs}}{
the full version~\cite{full-version}}}.
Let us simplify the result for a common case:
\begin{corollary}
  If $G$ is unweighted, $\alpha$ a constant and $\delta := 1/n$ to get high probability, 
  the (expected) running time of Algorithm~\ref{alg:approx-diag-omega}
  becomes $\tilde{\Oh}(m(\log^{1/2}n \log(n/\epsilon) + \epsilon^{-2} \log n))$.
  Assuming $\epsilon$ is small enough so that $\log n \leq 1/\epsilon$, we can further simplify this to $\tilde{\Oh}(m \epsilon^{-2} \log^{3/2}n)$.
\end{corollary}

This is nearly-linear in $m$, which is also true for the JLT-based approximation 
(with high probability) of Jin \etal~\cite{Zhang19}.
They state a running time of $\tilde{\Oh}(m \epsilon^{-2} \log^{5/2} n \log(1/\epsilon))$
for unweighted $G$ and fixed $\alpha = 1$. While we save at least a factor of $\log n$,
they achieve a relative approximation guarantee, which is difficult to compare to ours. 


\section{Group Forest Closeness Centrality}

\changed{
Since their introduction by Everett and Borgatti~\cite{everett99gc},
group centrality measures have been used in various
applications (see~\cite{Grinten2020ScalingUN}).
These measures indicate the importance of whole vertex
sets -- together as a group. They usually favor sets that
\enquote{cover} the graph well. Intuitively, a
group variant of forest closeness should reward vertex sets
that are ``forest-close'' to the remainder of the graph.
More formally,} to extend the concept of forest closeness
to groups of vertices, it is enough
to define the forest farness $\fdistu{S}$ of a set $S$ of vertices;
the forest closeness of $S$ is then
given by $\fdistinv{S} := \frac 1{\fdistu{S}}$.
Recall (from Proposition~\ref{forest-resistance})
that the forest farness of a single vertex $v$
of $G$ is identical to the electrical farness of $v$ in
the augmented graph $\Aug$.
We use this fact to generalize the forest farness
of a set $S$ of vertices of $G$. In particular, we
define $\fdistu{S} := \trace{(\msub{\AugLap}{S})^{-1}}$,
where $\AugLap$ is the Laplacian matrix of the augmented graph $\Aug$
and by $\msub{\AugLap}{S}$ we denote the matrix that is obtained from
$\AugLap$ by removing all rows and columns with indices in $S$.
This definition is based on a corresponding
definition of electrical farness by Li \etal~\cite{DBLP:conf/www/0002PSYZ19}.
For $|S| = 1$, it coincides with the definition of
electrical closeness from Section~\ref{sec:prelim}
~\cite{Izmailian_2013};
thus, our definition of group forest closeness
is compatible with the definition of the
forest closeness of individual vertices
(\ie Definition~\ref{forest-dist-vertex}).

Given our definition, it is natural to ask for
a set $S$ of $k$ vertices that maximizes
$\fdistinv{S}$ over all possible size-$k$ sets $S$;
indeed, this optimization problem has also
been considered for many other group centrality measures~\cite{Grinten2020ScalingUN}.
The following theorem
settles the complexity of the problem:

\begin{theorem}
\label{thm:GFC-NP-hard}
	Maximizing \textsc{GroupForestCloseness} subject to a cardinality constraint is $NP$-hard.
\end{theorem}

As Li \etal's~\cite{DBLP:conf/www/0002PSYZ19} hardness proof for group electrical closeness,
our reduction is from the vertex cover problem on 3-regular graphs.
Let $G = (V, E)$ be a 3-regular graph with $n$ vertices.
Our proof shows that there is a vertex cover of size $k$ in $G$ if and only if
the maximum group forest
closeness over all sets of size $k$ in $G$ exceeds a certain threshold.
We make use of the following property
that is adapted from a similar result by Li \etal:
\begin{lemma}
	\label{lemma:a-vc}
	Let $G$ be a connected and unweighted 3-regular graph and let $S \subset V$, $|S| = k \geq 1$.
	Then $\trace{(\msub{\Lap}{S}+\mat{I})^{-1}} \geq (n - k) / 4$
	and equality holds if and only if $S$ is a vertex cover of $G$.
\end{lemma}
Our proof of Theorem~\ref{thm:GFC-NP-hard}
exploits the fact that we can decompose
$\msub{\AugLap}{S}$ into a block that corresponds to the
universal vertex of $\Aug$ and into a block
that equals $\msub{\Lap}{S} + \mat{I}$.
This allows us to
apply the block-wise inversion and the Sherman-Morrison formula
to partially invert $\msub{\AugLap}{S}$.
In turn, we can apply
Lemma~\ref{lemma:a-vc} to bound
$\trace{(\msub{\AugLap}{S})^{-1}}$. The proof of
Lemma~\ref{lemma:a-vc} and the full proof
of Theorem~\ref{thm:GFC-NP-hard} can be
found in \changed{\arxivOrCamera{\Cref{apx:technical-proofs}}{the full version~\cite{full-version}}}.

Since an efficient algorithm for maximizing group forest closeness is unlikely to exist
(due to Theorem~\ref{thm:GFC-NP-hard}), 
it is desirable to construct an inexact algorithm
for this problem.
The next two results enable the construction of
such an algorithm;
they follow immediately
from respective results on group electrical closeness
on $\Aug$
(see Ref.~\cite[Theorem 5.4 and Theorem 6.1]{DBLP:conf/www/0002PSYZ19}).

\begin{lemma}
	$\fdistu{.}$ is a non-increasing
	and supermodular set function.
\end{lemma}

For the following corollary, we consider a greedy algorithm
that constructs a set $S$ of size $k$.
This set is initially empty; while $|S|$ is smaller than $k$,
the algorithm adds the vertex $v$ to $S$ that
maximizes the marginal gain: $v = \argmax_{x \in V \setminus S} \fdistu{S} - \fdistu{S \cup \{v\}}$.

\begin{corollary}
	The greedy algorithm computes a set $S$ such that:
	\[ \fdistu{\{v_0\}} - \fdistu{S} \geq \left(1 - \frac k{e(k - 1)}\right) \left(\fdistu{v_0} - \fdistu{\widetilde{S}}\right), \]
	where $v_0$ is the vertex with highest (individual) forest closeness
	and $\widetilde{S}$ is the set of size $k$ that maximizes group forest closeness.
\end{corollary}

\begin{algorithm}[tb]
\begin{algorithmic}[1]
	\State \textbf{Input:} Undir.\ graph $G = (V, E)$, group size $k$
	\State \textbf{Output:} Group $S \subseteq V$ of $k$ vertices
	\State $\mat{P} \gets \Call{pseudoInverse}{\AugLap}$
	\State $v \gets \argmin_{v \in V} n (\AugLap^\dagger[v, v]) + \trace{\mat{P}}$
	\State $\mat{M} \gets \Call{inverse}{\msub{\AugLap}{\{v\}}}$
		\Comment{Invariant: $\mat{M} \gets \msub{\AugLap}{S}^{-1}$ throughout the algorithm}
	\State $S \gets \{v\}$
	\While{$|S| \leq k$}
		\State $v \gets \argmax_{v \in V \setminus S} \frac{(\mat{M} e_v)^T (\mat{M} e_v)}{e_v^T \mat{M} e_v}$
		\State $\mat{M} \gets \msub{\mat{M}
			- \frac{\mat{M} e_v e_v^T  \mat{M}}{e_v^T \mat{M} e_v}}{\{v\}}$
			\label{line:inv-update}
		\State $S \gets S \cup \{v\}$
	\EndWhile
\end{algorithmic}
\caption{Greedy algorithm for group forest closeness maximization adapted from Li \etal}
\label{algo:li-greedy}
\end{algorithm}

Note that a na\"ive implementation of the greedy algorithm
would invert $\msub{\AugLap}{(S \cup \{v\})}$ for
each $v$, \ie it would require $k \cdot n$
matrix inversions in total.
By using the ideas of Li \etal for group electrical
closeness~\cite{DBLP:conf/www/0002PSYZ19}
(depicted in Algorithm~\ref{algo:li-greedy}
for the case of group forest closeness),
these inversions can be avoided, such
that only a single matrix inversion is required in total.
This makes use of the fact that whenever a vertex $u$
is added to the set $S$, we can decompose
$\msub{\AugLap}{S}$ into a block that consists
of $\msub{\AugLap}{(S \cup \{u\})}$ and
a single row/column that corresponds to $u$.
It is now possible to apply block-wise matrix inversion
to this decomposition to avoid the need to
recompute $(\msub{\AugLap}{(S \cup \{u\})})^{-1}$
from scratch (in line~\ref{line:inv-update} of the pseudocode).
We remark that the
greedy algorithm can be further accelerated
by utilizing the Johnson-Lindenstrauss
lemma~\cite{DBLP:conf/www/0002PSYZ19};
however, since this necessarily results in lower accuracy, we do not consider this extension in our experiments.

Furthermore, we note that by applying a standard reduction
by Gremban~\cite{GrembanPHD}, it would also be possible
to apply our UST-based algorithm (\ie Algorithm~\ref{alg:approx-diag-omega})
to the case of group forest closeness.
However, if the aforementioned
block-wise matrix inversion 
is not applied, this would require us to
sample USTs for each of the $k \cdot n$ vertex
evaluations.
On the other hand, in order to apply block-wise inversion,
the entire inverse of $\msub{\AugLap}{S}$ must be available (and not only the diagonal).
Computing this inverse via UST sampling is prohibitively expensive so far.
Hence, in our experiments, we prefer the algorithmic approach
by Li \etal (adapted for group forest closeness).

\section{Experiments}
\label{sec:experiments}

We study the empirical performance of our algorithms
on real-world graphs
and their impact on graph mining tasks.

\paragraph{Settings.}
Unless stated otherwise, all algorithms are implemented
in C++, using the NetworKit~\citep{DBLP:journals/netsci/StaudtSM16}
graph APIs. All experiments are conducted on
Intel Xeon Gold 6126 machines with $2 \times 12$ cores and 192 GiB of RAM each.
Unless stated otherwise, all experiments run on a
single core.
To ensure reproducibility, all experiments are managed by the
\textsc{SimExPal}~\citep{angriman2019guidelines} software.
For the evaluation, we use a large collection of undirected graphs
of different sizes, coming from a diverse set of domains.
All graphs have been downloaded from the public repositories
KONECT~\cite{kunegis13}, OpenStreetMap\footnote{\url{ https://www.openstreetmap.org}} 
and NetworkRepository~\cite{nr}.
We denote our proposed algorithm for forest closeness by \algo and set
$\alpha = 1$ (as done in Ref.~\cite{Zhang19}) 
in all experiments.

\paragraph{Competitors.}

\changed{For the forest closeness of individual vertices,
the main competitor is the JLT-based algorithm by Jin \etal~\cite{Zhang19},
which uses the Laplacian solver from Ref.~\cite{kyng16}.
We compare against two implementations of this algorithm;
one provided by the authors written in Julia v1.0.2
and our own implementation based on \textsc{Eigen}'s CG algorithm.\footnote{
\url{http://eigen.tuxfamily.org}.}
We denote them by \jltKyng and \jltCpp, respectively.}
Like in Ref.~\cite{Zhang19}, we compute the number
of linear sytems for \jltKyng and \jltCpp as
$\left\lceil \frac{\log n}{\epsilon^2} \right\rceil$
\changed{(which gives an $\epsilon \cdot c$ approximation for a fixed constant $c > 1$)}.

\subsection{Performance of \algo.}
\label{sec:ex-indiviual}

We measure now the performance
of \algo compared to the state-of-the art competitors.
Each method is executed with multiple settings
of its respective quality parameter.


\begin{figure}[tb]
\centering
\begin{subfigure}[t]{\columnwidth}
\centering
\includegraphics{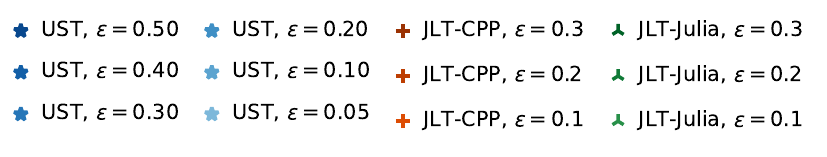}
\end{subfigure}

\begin{subfigure}[t]{.5\columnwidth}
\centering
\includegraphics{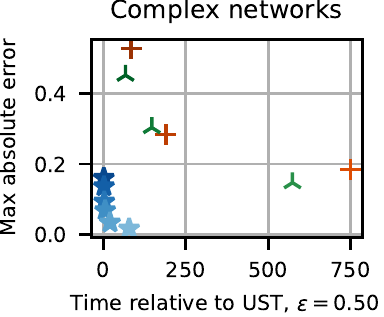}
\end{subfigure}\hfill
\begin{subfigure}[t]{.5\columnwidth}
\centering
\includegraphics{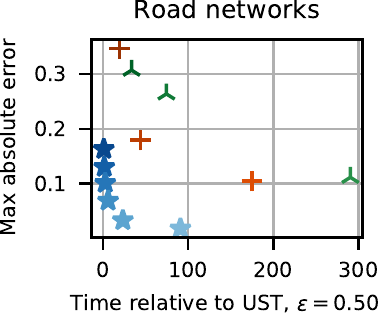}
\end{subfigure}

\caption{$\max_u |\Pomega[v, v] - \widetilde{\Pomega}[v, v]|$ over the instances
in Table~\ref{tab:time-kt}.}
\label{fig:quality-single-vertex}
\end{figure}

\paragraph{Accuracy and Running Time.}
We report the maximum absolute error
of the estimated diagonal values
(\ie $\max_v |\Pomega[v, v] - \widetilde{\Pomega}[v, v]|$)
over all vertices and instances from Table~\ref{tab:time-kt}.\footnote{Note that the top vertices in the forest closeness ranking
	are the ones with the \emph{lowest} $\Pomega[v, v]$ (see Eq.~\eqref{eq-fdistu-tr});
	hence, we also evaluate the ranking accuracy in a following
	experiment.}
As ground truth, we take $\Pomega[v, v]$ values
that are computed using Eigen's CG solver with a tolerance of $\lamgTol$;
exact inversion of $(\Lap + \mat{I})$ would
be infeasible for many of the input graphs.
A preliminary comparison against the values of $\Pomega[v, v]$ computed with the
NumPy \texttt{pinv} function demonstrated that CG provides a sufficiently
accurate ground truth.

Figure~\ref{fig:quality-single-vertex} shows that \algo achieves the best
results in terms of quality and running time for both complex and road
networks. More precisely, for complex networks and $\epsilon = 0.4$, \algo
yields a maximum absolute error of $\ustMaxAbsFour$, which is less than the
most accurate result of both competitors ($\jltJuliaMaxAbsOne$ achieved by
\jltKyng with $\epsilon = 0.1$), while being $\ustJuliaFourOneSpeedup$ faster.
Also, the running time of \algo does not increase sub\-stantially for
lower values of $\epsilon$, and its quality does not deteriorate quickly for
higher values of $\epsilon$. A similar pattern is observed for road networks as
well.


\setlength{\tabcolsep}{2pt}
\begin{table}[tb]
\centering
\footnotesize
\begin{tabular}{c}
Complex networks
\end{tabular}

\begin{tabular}{lrrrrrrr}
\toprule
\multirow{2}{*}{Graph} & \multirow{2}{*}{$|V|$} & \multirow{2}{*}{$|E|$} & \multicolumn{2}{c}{Time (s)} & \multicolumn{2}{c}{KT}\\
& & & UST & JLT & UST & JLT\\
\midrule
loc-brightkite\_edges & 58K & 214K& \textbf{\numprint{46.4}}& \numprint{186.4}& \textbf{\numprint{0.98}}& \numprint{0.95}\\
douban & 154K & 327K& \textbf{\numprint{80.8}}& \numprint{370.9}& \textbf{\numprint{0.71}}& \numprint{0.61}\\
soc-Epinions1 & 75K & 405K& \textbf{\numprint{55.5}}& \numprint{339.6}& \textbf{\numprint{0.95}}& \numprint{0.90}\\
slashdot-zoo & 79K & 467K& \textbf{\numprint{59.9}}& \numprint{412.3}& \textbf{\numprint{0.95}}& \numprint{0.92}\\
petster-cat-household & 105K & 494K& \textbf{\numprint{61.8}}& \numprint{372.1}& \textbf{\numprint{0.98}}& \numprint{0.92}\\
wikipedia\_link\_fy & 65K & 921K& \textbf{\numprint{58.2}}& \numprint{602.9}& \textbf{\numprint{0.98}}& \numprint{0.96}\\
loc-gowalla\_edges & 196K & 950K& \textbf{\numprint{230.9}}& \numprint{1215.5}& \textbf{\numprint{0.99}}& \numprint{0.97}\\
wikipedia\_link\_an & 56K & 1.1M& \textbf{\numprint{50.7}}& \numprint{562.6}& \textbf{\numprint{0.96}}& \numprint{0.93}\\
wikipedia\_link\_ga & 55K & 1.2M& \textbf{\numprint{44.8}}& \numprint{578.6}& \textbf{\numprint{0.98}}& \numprint{0.97}\\
petster-dog-household & 260K & 2.1M& \textbf{\numprint{359.6}}& \numprint{2472.1}& \textbf{\numprint{0.98}}& \numprint{0.96}\\
livemocha & 104K & 2.2M& \textbf{\numprint{107.4}}& \numprint{1429.3}& \textbf{\numprint{0.98}}& \numprint{0.97}\\
\bottomrule
\end{tabular}

\smallskip

\begin{tabular}{c}
Road networks
\end{tabular}

\begin{tabular}{lrrrrrrr}
\toprule
\multirow{2}{*}{Graph} & \multirow{2}{*}{$|V|$} & \multirow{2}{*}{$|E|$} & \multicolumn{2}{c}{Time (s)} & \multicolumn{2}{c}{KT}\\
& & & UST & JLT & UST & JLT\\
\midrule
mauritania & 102K & 150K& \textbf{\numprint{98.1}}& \numprint{217.6}& \textbf{\numprint{0.88}}& \numprint{0.77}\\
turkmenistan & 125K & 165K& \textbf{\numprint{118.5}}& \numprint{273.6}& \textbf{\numprint{0.92}}& \numprint{0.85}\\
cyprus & 151K & 189K& \textbf{\numprint{149.4}}& \numprint{315.8}& \textbf{\numprint{0.89}}& \numprint{0.80}\\
canary-islands & 169K & 208K& \textbf{\numprint{185.5}}& \numprint{382.0}& \textbf{\numprint{0.92}}& \numprint{0.84}\\
albania & 196K & 223K& \textbf{\numprint{192.6}}& \numprint{430.2}& \textbf{\numprint{0.90}}& \numprint{0.82}\\
benin & 177K & 234K& \textbf{\numprint{188.1}}& \numprint{406.8}& \textbf{\numprint{0.92}}& \numprint{0.83}\\
georgia & 262K & 319K& \textbf{\numprint{322.1}}& \numprint{605.3}& \textbf{\numprint{0.91}}& \numprint{0.83}\\
latvia & 275K & 323K& \textbf{\numprint{355.2}}& \numprint{665.4}& \textbf{\numprint{0.91}}& \numprint{0.83}\\
somalia & 291K & 409K& \textbf{\numprint{420.1}}& \numprint{747.5}& \textbf{\numprint{0.92}}& \numprint{0.84}\\
ethiopia & 443K & 607K& \textbf{\numprint{825.9}}& \numprint{1209.7}& \textbf{\numprint{0.91}}& \numprint{0.83}\\
tunisia & 568K & 766K& \textbf{\numprint{1200.1}}& \numprint{1629.0}& \textbf{\numprint{0.89}}& \numprint{0.79}\\
\bottomrule
\end{tabular}

\caption{Running time and KT ranking scores of \algo and JLT-based algorithms.
In the JLT column we report, for each instance, the competitor with highest KT
score. For equal KT scores (up to the second decimal place) we choose the
fastest competitor.}
\label{tab:time-kt}
\end{table}

\paragraph{Vertex Ranking.}
Moreover, we measure the accuracy in terms of vertex rankings,
which is often more relevant than individual scores~\citep{newman2018networks,okamoto2008ranking}.
In Table~\ref{tab:time-kt} we report
the Kendall's rank correlation coefficient (KT) of the vertex ranking \wrt\ 
the ground truth along with running times for complex and road networks.
For each instance, we pick the best \changed{run, \ie the UST and JLT columns display the run
with highest respective KT value.}
If the values are the same up to the second decimal place, we pick the fastest one.
\algo has consistently the best vertex ranking scores; at the same time, it is faster than the competitors.
In particular, \algo is on average $\ustJLTSpeedupCplx$ faster than the JLT-based approaches
on complex networks and $\ustJLTSpeedupRoad$ faster on road networks.

\paragraph{Parallel Scalability.}

\algo is well-suited for parallel implementations since
each UST can be sampled independently in parallel.
Hence, we  provide parallel implementations of \algo
based on OpenMP (for multi-core parallelism) and MPI
(to scale to multiple compute nodes).
The OpenMP implementation on 24 cores exhibits
a speedup of $\parSpeedupCplx$ on complex networks and
$\parSpeedupRoad$ on road networks --
more detailed results can be found in
\changed{\arxivOrCamera{\Cref{fig:par-scalability}, \Cref{apx:additional-exp}}{Figure 5 in the full
version~\cite{full-version}}}.
The results for MPI are depicted in \changed{\arxivOrCamera{\Cref{fig:distr-scalability},
\Cref{apx:additional-exp}}{Figure 3 in
the full version~\cite{full-version}}}.
In this setting, \algo obtains a speedup
of $\distrSpeedupCplx$ on complex and $\distrSpeedupRoad$ on road networks
on up to 16 compute nodes -- for this experiment we set $\epsilon = 0.1$ and
we use the instances in
\changed{\arxivOrCamera{\Cref{tab:large}, \Cref{apx:additional-exp}}{Table 2 in
the full version~\cite{full-version}}}.
More sophisticated load balancing
techniques are likely to increase the speedups
in the MPI setting; they are left for future work.
Still, the MPI-based algorithm can rank complex networks with up to $334$M edges in
less than $20$ minutes. Road networks with $31$M edges take less than $25$ minutes.

\subsection{Semi-Supervised Vertex Classification.}
\label{sec:ex-group}

To demonstrate the relevance of \changed{group forest closeness} in
graph mining applications, we
apply them to semi-supervised vertex classification~\cite{chapelleSZ09}. Given
a graph $G$ with labelled vertices, the goal is to predict the labels of all
vertices of $G$ by training a classifier using a small set of labelled vertices
as training set. The choice of the vertices for the training set can influence
the accuracy of the classifier, especially when the number of labelled vertices
is small compared to $|V|$~\citep{oleks2018, Avrachenkov2013}.

A key aspect in semi-supervised learning problems is the so-called
\emph{cluster assumption} \ie vertices that are close or that belong to the
same cluster typically have the same label~\cite{zhouBLWS03,chapelleWS02}.
Several models label vertices by propagating information through the
graph via diffusion~\cite{chapelleSZ09}.
\changed{We expect group forest closeness to cover the graph
more thoroughly than individual forest closeness.
Hence, we conjecture that} choosing vertices
with high group centrality improves diffusion and thus the accuracy of
propagation-based models.
We test this hypothesis by
comparing the classification accuracy of the label
propagation model~\cite{chapelleSZ09, zhouBLWS03}
where the training set is chosen using different strategies.\footnote{While this
model is less powerful than state-of-the-art predictors,
our strategy to select the training set could also
be applied to more sophisticated models like graph neural networks.}
The main idea of label propagation is to start from a small number of labelled
vertices and each vertex iteratively propagates its label to its neighbors
until convergence.

In our experiments, we use the Normalized Laplacian variant of label
propagation~\cite{zhouBLWS03}. We set the return probability hyper-parameter to
$0.85$, and we evaluate its accuracy on two well-known disconnected graph
datasets: Cora ($|V| = \numprint{2708}, |E| = \numprint{5278}$) and
Citeseer($|V| = \numprint{3264}, |E| = \numprint{4536}$)~\cite{senNBGGE08}.
Since this variant of label propagation cannot handle graphs with isolated
vertices (\ie zero-degree vertices), we remove all isolated vertices from these datasets.
For a fixed size $k$ of the training set, we select its vertices as the group of
vertices computed by our greedy algorithm for group forest
maximization and as the top-$k$ vertices with highest estimated forest
closeness. We \changed{include several well-known (individual) vertex selection
strategies for comparison}:
average over 10 random trials, the top-$k$ vertices with highest degree, the
top-$k$ vertices with highest betweenness centrality and the top-$k$ vertices
with highest Personalized PageRank.

\begin{figure}
\centering
\begin{subfigure}[t]{\columnwidth}
\centering
\includegraphics{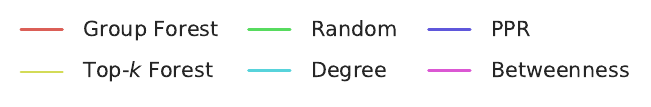}
\end{subfigure}

\begin{subfigure}[t]{.5\columnwidth}
\centering
\includegraphics{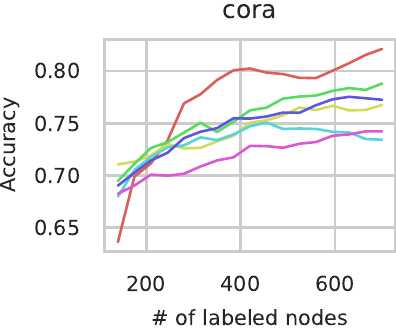}
\end{subfigure}\hfill
\begin{subfigure}[t]{.5\columnwidth}
\centering
\includegraphics{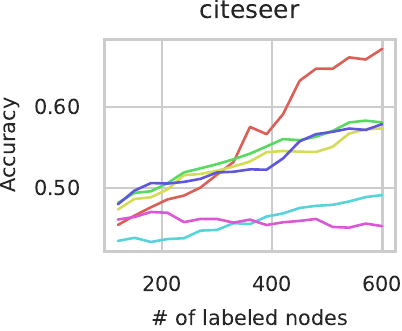}
\end{subfigure}
\caption{Accuracy in semi-supervised vertex classification
when using different strategies to create the training set.}
\label{fig:vertex-class}
\end{figure}

Figure~\ref{fig:vertex-class} shows that on
graphs with disconnected components and for a moderate number of
labelled vertices, selecting the training
set by group forest closeness maximization yields consistently
superior accuracy than strategies based on existing
centrality measures (including top-$k$ forest closeness).
As expected, the accuracy of existing measures
improves if one considers connected graphs
(\changed{\arxivOrCamera{\Cref{fig:vertex-class-lcc}, \Cref{apx:additional-exp}}{Figure 6 in the full version~\cite{full-version}}});
yet, group forest closeness is nearly as accurate as the best competitors
on these graphs.

\changed{The running time of our greedy algorithm for group forest maximization
is reported in \arxivOrCamera{\Cref{tab:time-group}, \Cref{apx:additional-exp}}{the full version~\cite{full-version}}}.

\section{Conclusions}
\label{sec:conclusions}
In this paper, we proposed a new algorithm to approximate
forest closeness faster and more accurately than previously possible.
We also generalized the definition of forest closeness
to group forest closeness and demonstrated that
for semi-supervised vertex classification in disconnected
graphs, group forest closeness outperforms existing
approaches.
In future work, we want to consider extensions of our approaches
to directed graphs. Another challenging
extension would involve generalizing an approach
based on USTs to group forest closeness
to improve upon the performance of our greedy algorithm.

\ifblind
\else
\section{Acknowledgments}
\begin{acks}
TODO: here only people, grants on the first page with thanks
\end{acks}
\fi

\bibliographystyle{abbrv}
{\footnotesize 
\itemsep1em
\bibliography{references}
}

\arxivOrCamera{
\clearpage
\appendix
\section{Technical Proofs}
\label{apx:technical-proofs}

\begin{proof}(Proposition~\ref{prop:unbiased})
Since $\{\pivot,v\} \in E'$, we have in the unweighted case that $\effresG{\Aug}{\pivot}{v}$ is the number of spanning
trees of $\Aug$ that contain $\{\pivot,v\}$ divided by the number of all spanning trees of
$\Aug$ (follows from Kirchhoff's theorem, see~\cite[Ch.~II]{DBLP:books/daglib/0009415}).
In the weighted case, replace ``number'' by ``total weight'', respectively
(where the weight of a UST is the product of all edge weights).
We focus on the unweighted case in the following for ease of exposition; the proof for the weighted case works
in the same way.

Clearly, $R[v] / \tau$, as used by Algorithm~\ref{alg:approx-diag-omega}, is an estimator for
$\effresG{\Aug}{\pivot}{v}$. It remains to show that it is unbiased, \ie $\E[R[v]/\tau] = \effresG{\Aug}{\pivot}{v}$.
To this end, let $T_i$ be the UST sampled in iteration $i$ and $X_{i,v}$ the random indicator variable with
$X_{i,v} = 1$ if $\{\pivot,v\} \in T_i$ and $0$ otherwise. Then:
\begin{align*}
	\E[R[v]/\tau] &= \frac{1}{\tau} \E[R[v]]
		= \frac{1}{\tau} \sum_{i=1}^\tau \mathbb{P}[\{\pivot,v\} \in T_i] \cdot X_{i,v} \\
	& = \frac{1}{\tau} \sum_{i=1}^\tau \effresG{\Aug}{\pivot}{v} \cdot 1
		= \effresG{\Aug}{\pivot}{v},
\end{align*}
which follows from the definition of expectation and the above correspondence between
(the relative fre\-quency of an edge in) USTs and effective resistances.
\end{proof}

\begin{proof}(Proposition~\ref{prop:ust-wilson-time})
By plugging the augmented graph $\Aug$ (with constant diameter)
into the proof of Lemma~10 of Ref.~\cite{angrimanPGM20}, we obtain for the running time $W(n)$
on a graph with $n$ vertices: $W(n) = \Oh(\operatorname{vol}(\Aug)) = \Oh(\alpha \operatorname{vol}(G) + n)$ expected time per call
in Line~\ref{line:ust-sampling}.\hfill
\end{proof}

\begin{proof}(Theorem~\ref{thm:time-approx})
For the linear system in Line~\ref{line:linear-system}, we employ the
SDD solver by Cohen \etal~\cite{CohenKyng14}; it takes $\tilde{\Oh}(m \log^{1/2} n \log(1/\eta))$
time to achieve a relative error bound of $\Vert \myvec{\tilde{x}} - \myvec{x} \Vert_{\mat{L'}}$, where $\mat{L'} := \alpha \mat{L} + \mat{I}$.
We can express the equivalence of this matrix-based norm with the maximum norm by
adapting Lemma~12 of Ref.~\cite{angrimanPGM20} with the norm for $\mat{L'}$ (instead of $\mat{L}$):
$\sqrt{\mu_1} \cdot \Vert \myvec{x} \Vert_\infty \leq \Vert \myvec{x} \Vert_{\mat{L'}} \leq \sqrt{\alpha (c+2) \operatorname{vol}(G)} \Vert \myvec{x} \Vert_\infty$,
where $\mu_1$ is the smallest eigenvalue of $\mat{L'}$. In fact, $\mu_1 = \alpha \lambda_1 + 1 = 1$,
where $\lambda_1 = 0$ is the smallest eigenvalue of $\mat{L}$, so that we can simplify: 
\begin{equation}
\Vert \myvec{x} \Vert_\infty \leq \Vert \myvec{x} \Vert_{\mat{L'}} \leq \sqrt{\alpha (c+2) \operatorname{vol}(G)} \Vert \myvec{x} \Vert_\infty.
\end{equation}

Let us set $c := \frac{n}{\alpha \cdot \operatorname{vol}(G)}$; by our assumption in the theorem,
$c$ is a constant. Hence, if we set $\eta := \kappa \epsilon / 6 \sqrt{\alpha (c+2) \operatorname{vol}(G)}$,
the SDD solver's accuracy can be bounded by:
\begin{align*}
  \Vert \myvec{\tilde{x}} - \myvec{x} \Vert_\infty & \leq \Vert \myvec{\tilde{x}} - \myvec{x} \Vert_{\mat{L'}} \leq \eta \cdot \Vert \myvec{x} \Vert_\mat{L'} \\
  & \leq \eta \sqrt{\alpha (c+2) \operatorname{vol}(G)} \Vert \myvec{x} \Vert_\infty \\
  & = \frac{\kappa \epsilon}{6} \Vert \myvec{x} \Vert_\infty \leq \frac{\kappa \epsilon}{3}.
\end{align*}

The last inequality follows from the fact that the values in $\myvec{x}$ are bounded by the effective
resistance, which in turn is bounded by the graph distance and thus $2$ (via the edges to/from $u$).
If each entry has accuracy of $\kappa \epsilon / 3$ (or better),
then Eq.~(\ref{eq:forest-dist-pair}) is solved with accuracy $\kappa \epsilon$ (or better).
The resulting running time for the SDD solver is thus $\tilde{\Oh}(m \log^{1/2} n \log(1 / \eta))
= \tilde{\Oh}(m \log^{1/2} n \log(\sqrt{\alpha \operatorname{vol}(G)} / \epsilon))$.

According to Proposition~\ref{prop:ust-wilson-time} and with $n \leq c \cdot \alpha \cdot \operatorname{vol}(G)$,
sampling one UST takes $\Oh(\alpha \operatorname{vol}(G))$ expected time. It remains to identify
a suitable sample size $\tau$ for the approximation to hold. To this end, let $\epsilon' := (1-\kappa)\epsilon$
denote the tolerable absolute error for the UST-based approximation part.
Plugging $\tau := \lceil \log(2m/\delta) / 2(\epsilon')^2\rceil$ into the proof of Theorem~3 of Ref.~\cite{angrimanPGM20}
(and thus essentially Hoeffding's bound) with the fact that the eccentricity of $u$
is $1$, we obtain the desired result.\hfill
\end{proof}

\begin{proof}(Lemma~\ref{lemma:a-vc})
The proof in Li \etal~\cite[Lemma~4.1]{DBLP:conf/www/0002PSYZ19} exploits (among others) that the diagonal is constant.
If we replace $3$ by $4$, this argument and all others (such as positive definiteness)
still hold and the result becomes $(n-k)/4$ instead of $(n-k)/3$.\hfill
\end{proof}

\begin{proof}(Theorem~\ref{thm:GFC-NP-hard})
	Let $G$ be 3-regular and let $S \subset V$, $|S| = k$.
	We prove that $f(S) \geq \frac{4}{3n+k} + (\frac{1}{4} + \frac{1}{4(3n+k)}) (n-k) =: t(n,k)$, where equality
	holds if and only if $S$ is a vertex cover of $G$.
	Let $\mat{A}$ be the $(n-k) \times (n-k)$ submatrix of $\msub{\AugLap}{S}$ that corresponds
	to all vertices except the universal vertex, \ie $\mat{A} := \msub{\Lap}{S} + \mat{I}$.
	Note that $\mat{A}$ is symmetric.
	Since $G$ is 3-regular, all diagonal entries
	of $\mat{A}$ are 4. All non-diagonal entries
	have value $-1$ and there can be
	at most three such entries per row / column of $\mat{A}$.
	In particular, the row and column sums of $\mat{A}$
	are all $\geq 1$.
	An elementary calculation (\ie expanding the $ij$-th
	element of the matrix multiplication $A$ times $A^{-1}$, and
	summing over $j$) shows:

	\begin{equation}
		\label{eq:rowcol-relation}
		\left(\sum_{\ell} \mat{A}_{i \ell}\right) \left(\sum_{\ell} \mat{A}^{-1}_{\ell i}\right) = 1,
	\end{equation}
	hence the row sums and column sums of $\mat{A}^{-1}$
	are all $\leq 1$.
	Let us now decompose $\msub{\AugLap}{S}$ into blocks as follows:
	\[
		\msub{\AugLap}{S} = \left(\begin{array}{c|cccc}
			n & -1 & \ldots & -1 \\ \hline
			-1 & & & \\
			\ldots & & \mat{A} & \\
			-1 & & & \\
		\end{array}\right).
	\]
	By blockwise inversion we obtain:
	\[
		(\msub{\AugLap}{S})^{-1} = \left(\begin{array}{c|cccc}
			\frac{1}{n - \myvec{1}^T \mat{A}^{-1} \myvec{1}}  &  & \ldots & \\ \hline
			& & & \\
			\ldots & & (\mat{A} - \frac 1n \mat{J})^{-1} & \\
			& & & \\
		\end{array}\right),
	\]
	where $\mat{J}$ is the $(n - k) \times (n - k)$ matrix of all ones.
	To compute $(\mat{A} - \frac{1}{n} \mat{J})^{-1}$, we notice that
	$-\frac{1}{n}\mat{J}$ can be written as $\myvec{1}^T \cdot (-1 \frac 1n) \myvec{1}$
	and apply the Sherman-Morrison formula. This yields
	\begin{equation}
		\label{eq:ablock-inverse}
		(\mat{A} - \frac{1}{n} \mat{J})^{-1} = \mat{A}^{-1} + \frac{1}{n - \myvec{1}^T \mat{A}^{-1} \myvec{1}} \mat{A}^{-1} \mat{J} \mat{A}^{-1}.
	\end{equation}
	We note that $\myvec{1}^T \mat{A}^{-1} \myvec{1}$ is equal to
	the sum of all entries of $\mat{A}^{-1}$ and this is bounded by
	the sum of all column sums of $\mat{A}^{-1}$, \ie
  	$\myvec{1}^T \mat{A}^{-1} \myvec{1} \leq n - k < n$
	and the denominator of Eq.~(\ref{eq:ablock-inverse}) is well-defined.
	Also, we have $\trace{\mat{A}^{-1} \mat{J} \mat{A}^{-1}} = \sum_{v \in V \setminus S} (\sum_j \mat{A}^{-1}_{vj}) (\sum_i \mat{A}^{-1}_{iv})$
	and thus $\trace{(\mat{A}^{-1} - \frac 1n \mat{J})^{-1}}$ only depends on $\trace{\mat{A}^{-1}}$
	and row/column sums of $\mat{A}^{-1}$.

	Now consider the case that $S$ is a vertex cover.
	In this case, $\mat{A}$ has no off-diagonal entries
	(and all row (or column) sums of $\mat{A}$ are 4).
	For the entry $(\msub{\AugLap}{S})^{-1}[1][1]$, we then obtain using Lemma~\ref{lemma:a-vc}:
	$1/(n-(n-k)/4) = 4/(3n+k)$. The inverse $(\mat{A} - \frac 1n \mat{J})^{-1}$, in turn,
	resolves to $\frac{1}{4} (\mat{I} + \frac{1}{3n+k} \mat{J})$, so that we obtain
	$\trace{(\msub{\AugLap}{S})^{-1}} = t(n,k)$.

	On the other hand, assume that $S$ is not a vertex cover.
	In this case, $\mat{A}$ is entry-wise smaller than or equal to
	the vertex cover case.
	Furthermore, at least one element is now strictly smaller,
	\ie there exists rows/columns of $\mat{A}$
	whose sum is smaller than 4. Due to Eq.~(\ref{eq:rowcol-relation}),
	this implies that some row/column sums of $\mat{A}^{-1}$
	are strictly larger than in the vertex cover case
	(namely, the rows/columns of $\mat{A}$ that sum to less than 4)
	and all others are equal to the vertex cover case
	(\ie the rows/columns of $\mat{A}$ that still sum to 4).
	Furthermore, by applying Lemma~\ref{lemma:a-vc},
	we notice that $\trace{\mat{A}^{-1}}$ is now larger
	compared to the vertex cover case.
	Since $\trace{(\mat{A} - \frac{1}{n} \mat{J})^{-1}}$ only depends
	on $\trace{\mat{A}^{-1}}$ and the row/column sums of $\mat{A}^{-1}$,
	the final trace can only be strictly larger than in the vertex cover case.

	\hfill
\end{proof}


\section{Algorithmic Details}
\label{sec:app-algorithmic-details}

\begin{algorithm}[H]
  \begin{algorithmic}[1]
    \begin{small}
    \Function{SamplingUST}{$G$, $\pivot$}
    \State \textbf{Input:} graph $G=(V,E)$, universal vertex $\pivot \in V$
    \State \textbf{Output:} $R :=$ estimated effective resistance values
    \State $R[v] \gets 0 ~\forall v \in V$ \label{line:start-init}
    \State $T \gets \{\pivot\}$
    \State Let $v_1, \ldots, v_n$ be a reordering of $V$ according to ascending degree
    \For{$i \gets 1$ to $n$}
    \State $P \gets $ random walk on $G$ from $v_i$ to $T$
    \State $LE(P) \gets$ loop erasure of $P$ in order of appearance
    \State $T \gets T \cup LE(P)$
    \If{ last vertex of $LE(P) = \pivot$}
    \State $w \gets$ last visited vertex before $\pivot$
    \State $R[w] \gets R[w] + 1$
    \label{line:effres-update}
        \EndIf
    \EndFor \label{line:end-sampling}
    \State \textbf{return} $R$
    \EndFunction
    \end{small}
  \end{algorithmic}
  \caption{Sampling algorithm for USTs (based on Wilson's algorithm)}
  \label{alg:sampling-ust}
\end{algorithm}


\section{Additional Experimental Results}
\label{apx:additional-exp}
\paragraph{Average Accuracy.} To confirm that
\algo performs well on average and not only when considering
the maximal error over many instances, we additionally report the
\emph{average} (over all instances from Table~\ref{tab:time-kt}) of the absolute error
in Figure~\ref{fig:avg-abs-err}.

\begin{figure}[tb]
\centering
\begin{subfigure}[t]{.5\columnwidth}
\centering
\includegraphics{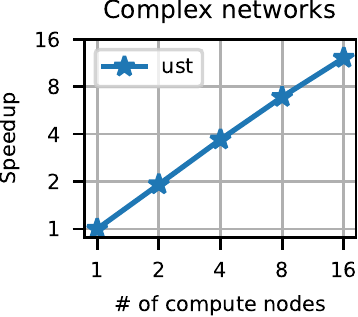}
\end{subfigure}\hfill
\begin{subfigure}[t]{.5\columnwidth}
\centering
\includegraphics{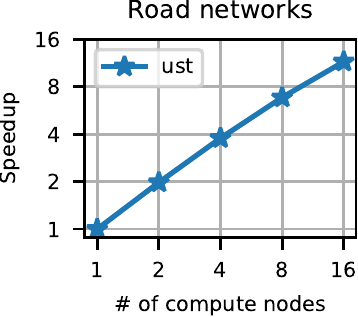}
\end{subfigure}
\caption{Geometric mean of the speedup of \algo with $\epsilon =
\numprint{0.1}$ on multiple compute nodes over a single compute node ($1
\times 24$ cores). Data points are aggregated over the instances in
\Cref{tab:large}.}
\label{fig:distr-scalability}
\end{figure}

\paragraph{Parallel Scalability.} In Figure~\ref{fig:par-scalability} we report the parallel scalability of \algo
on multiple cores. We hypothesize that the moderate speedup is mainly due to memory latencies:
while sampling a UST, our algorithm performs several random accesses to the graph data structure
(\ie an adjacency array), which are prone to cache misses.

Furthermore, Table~\ref{tab:large} reports detailed statistics about the instances used for
experiments in distributed memory along with running times of \algo on $16\times 24$ cores
with $\epsilon = 0.1$ and $\epsilon = 0.3$.

\paragraph{Vertex Classification.} Figure~\ref{fig:vertex-class-lcc} shows the accuracy in semi-supervised vertex
classification in connected graphs when using different strategies to create the training set.
Compared to disconnected graphs, the competitors perform better in this setting.
However, as described in Section~\ref{sec:ex-group}, choosing the training set by group forest
closeness maximization yields nearly the same accuracy as the best competitors in
our datasets.

\begin{figure}[tb]
\centering
\begin{subfigure}[t]{\columnwidth}
\centering
\includegraphics{./plots/legend-quality}
\end{subfigure}

\begin{subfigure}[t]{.5\columnwidth}
\centering
\includegraphics{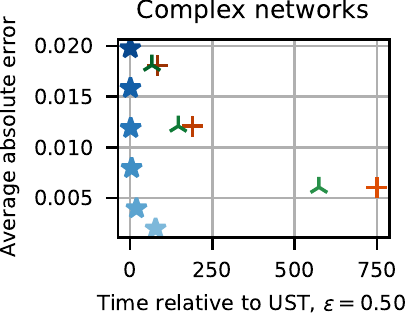}
\end{subfigure}\hfill
\begin{subfigure}[t]{.5\columnwidth}
\centering
\includegraphics{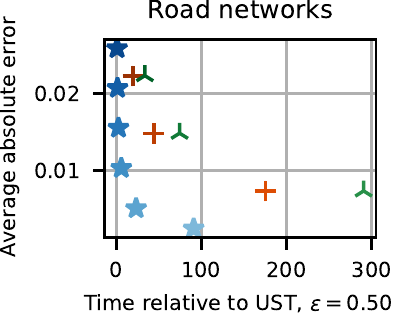}
\end{subfigure}
\caption{Arithmetic mean of the absolute errors $|\max_v \Pomega[v, v] -
\widetilde{\Pomega}[v, v]|$ over the instances in Table~\ref{tab:time-kt}.}
\label{fig:avg-abs-err}
\end{figure}

\begin{figure}[tb]
\centering
\begin{subfigure}[t]{.5\columnwidth}
\centering
\includegraphics{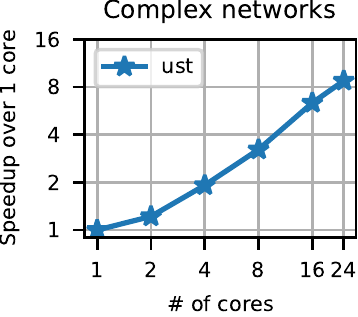}
\end{subfigure}\hfill
\begin{subfigure}[t]{.5\columnwidth}
\centering
\includegraphics{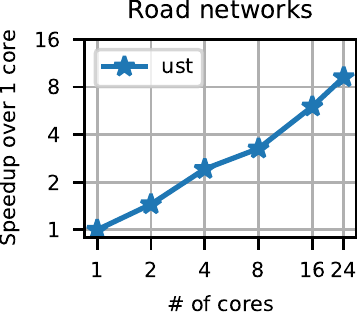}
\end{subfigure}
\caption{Geometric mean of the speedup of \algo with $\epsilon =
\numprint{0.05}$ on multiple cores over a sequential run (shared memory).
Data points are aggregated over the instances in Table~\ref{tab:time-kt}.}
\label{fig:par-scalability}
\end{figure}

\begin{figure}[p]
\centering
\begin{subfigure}[t]{\columnwidth}
\centering
\includegraphics{plots/legend-node-class}
\end{subfigure}

\begin{subfigure}[t]{.5\columnwidth}
\centering
\includegraphics{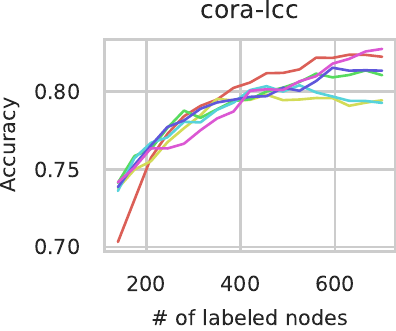}
\end{subfigure}\hfill
\begin{subfigure}[t]{.5\columnwidth}
\centering
\includegraphics{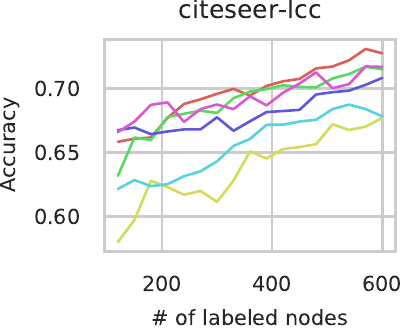}
\end{subfigure}
\caption{Accuracy in semi-supervised vertex classification on the largest
connected component of the datasets when using different strategies to create
the training set. Cora-lcc: $|V| = \numprint{2485}, |E| = \numprint{5069}$,
Citeseer-lcc: $|V| = \numprint{2110}, |E| = \numprint{3668}$.}
\label{fig:vertex-class-lcc}
\end{figure}

\begin{table}[p]
\centering\footnotesize
\begin{tabular}{c}
Complex networks
\end{tabular}

\begin{tabular}{lrrrr}
\toprule
\multirow{2}{*}{Graph} & \multirow{2}{*}{$|V|$} & \multirow{2}{*}{$|E|$} & \multicolumn{2}{c}{Time (s)} \\
 & & & $\varepsilon = 0.1$ & $\varepsilon = 0.3$\\
\midrule
soc-LiveJournal1 & \numprint{4846609} & \numprint{42851237} & \numprint{348.9} & \numprint{118.5}\\
wikipedia\_link\_fr & \numprint{3333397} & \numprint{100461905} & \numprint{205.4} & \numprint{90.7}\\
orkut-links & \numprint{3072441} & \numprint{117184899} & \numprint{293.5} & \numprint{92.2}\\
dimacs10-uk-2002 & \numprint{18483186} & \numprint{261787258} & \numprint{1101.3} & \numprint{365.8}\\
wikipedia\_link\_en & \numprint{13593032} & \numprint{334591525} & \numprint{919.3} & \numprint{295.4}\\
\bottomrule
\end{tabular}
\medskip

\begin{tabular}{c}
Road networks
\end{tabular}
\begin{tabular}{lrrrr}
\toprule
\multirow{2}{*}{Graph} & \multirow{2}{*}{$|V|$} & \multirow{2}{*}{$|E|$} & \multicolumn{2}{c}{Time (s)} \\
 & & & $\varepsilon = 0.1$ & $\varepsilon = 0.3$\\
\midrule
slovakia & \numprint{543733} & \numprint{638114} & \numprint{28.1} & \numprint{9.9}\\
netherlands & \numprint{1437177} & \numprint{1737377} & \numprint{82.9} & \numprint{31.1}\\
greece & \numprint{1466727} & \numprint{1873857} & \numprint{74.5} & \numprint{29.8}\\
spain & \numprint{4557386} & \numprint{5905365} & \numprint{273.0} & \numprint{86.2}\\
great-britain & \numprint{7108301} & \numprint{8358289} & \numprint{419.0} & \numprint{136.6}\\
dach & \numprint{20207259} & \numprint{25398909} & \numprint{1430.1} & \numprint{473.7}\\
africa & \numprint{23975266} & \numprint{31044959} & \numprint{1493.4} & \numprint{499.3}\\
\bottomrule
\end{tabular}

\caption{Large networks used for scalability experiments
in distributed memory and running time of \algo on $16\times 24$ cores.}
\label{tab:large}
\end{table}

\begin{table}[p]
\centering\footnotesize
\begin{tabular}{lrr}
\toprule
Graph & Group size & Time (s)\\
\midrule
\multirow{3}{*}{cora} & \numprint{200} & \numprint{1559.3}\\
 & \numprint{400} & \numprint{2210.6}\\
 & \numprint{600} & \numprint{2663.4}\\
\midrule
\multirow{3}{*}{citeseer} & \numprint{200} & \numprint{2518.6}\\
 & \numprint{400} & \numprint{3666.5}\\
 & \numprint{600} & \numprint{4642.4}\\
\bottomrule
\end{tabular}

\caption{Running time of our greedy algorithm for group forest maximization.}
\label{tab:time-group}
\end{table}

\begin{table}[p]
\centering\footnotesize
\begin{tabular}{c}
Complex networks
\end{tabular}

\begin{tabular}{lrrrrrr}\toprule
\multirow{2}{*}{Graph} & \multicolumn{6}{c}{Time (s)}\\
$\hfill\epsilon$& $0.05$& $0.1$& $0.2$& $0.3$& $0.4$& $0.5$\\
\midrule
loc-brightkite\_edges & \numprint{46.4} & \numprint{11.6} & \numprint{3.0} & \numprint{1.4} & \numprint{0.8} & \numprint{0.5}\\
douban & \numprint{80.8} & \numprint{20.5} & \numprint{5.2} & \numprint{2.4} & \numprint{1.5} & \numprint{0.9}\\
soc-Epinions1 & \numprint{55.5} & \numprint{14.0} & \numprint{3.5} & \numprint{1.6} & \numprint{1.0} & \numprint{0.7}\\
slashdot-zoo & \numprint{59.9} & \numprint{15.6} & \numprint{3.8} & \numprint{1.8} & \numprint{1.1} & \numprint{0.7}\\
petster-cat-household & \numprint{61.8} & \numprint{15.7} & \numprint{4.0} & \numprint{1.8} & \numprint{1.1} & \numprint{0.8}\\
wikipedia\_link\_fy & \numprint{58.2} & \numprint{15.0} & \numprint{3.9} & \numprint{1.9} & \numprint{1.1} & \numprint{0.8}\\
loc-gowalla\_edges & \numprint{230.9} & \numprint{63.0} & \numprint{15.7} & \numprint{7.1} & \numprint{4.4} & \numprint{2.8}\\
wikipedia\_link\_an & \numprint{50.7} & \numprint{12.1} & \numprint{3.1} & \numprint{1.5} & \numprint{0.9} & \numprint{0.7}\\
wikipedia\_link\_ga & \numprint{44.8} & \numprint{11.3} & \numprint{3.1} & \numprint{1.6} & \numprint{1.1} & \numprint{0.8}\\
petster-dog-household & \numprint{359.6} & \numprint{87.7} & \numprint{22.5} & \numprint{10.3} & \numprint{6.0} & \numprint{4.1}\\
livemocha & \numprint{107.4} & \numprint{28.6} & \numprint{7.3} & \numprint{3.5} & \numprint{2.1} & \numprint{1.5}\\
\bottomrule
\end{tabular}
\medskip

\begin{tabular}{c}
Road networks
\end{tabular}

\begin{tabular}{lrrrrrr}\toprule
\multirow{2}{*}{Graph} & \multicolumn{6}{c}{Time (s)}\\
$\hfill\epsilon$& $0.05$& $0.1$& $0.2$& $0.3$& $0.4$& $0.5$\\
\midrule
mauritania & \numprint{98.1} & \numprint{24.4} & \numprint{6.9} & \numprint{2.8} & \numprint{1.6} & \numprint{1.0}\\
turkmenistan & \numprint{118.5} & \numprint{30.2} & \numprint{7.7} & \numprint{3.4} & \numprint{2.1} & \numprint{1.3}\\
cyprus & \numprint{149.4} & \numprint{37.7} & \numprint{9.8} & \numprint{4.4} & \numprint{2.6} & \numprint{1.7}\\
canary-islands & \numprint{185.5} & \numprint{46.7} & \numprint{11.4} & \numprint{5.2} & \numprint{3.0} & \numprint{2.0}\\
albania & \numprint{192.6} & \numprint{52.6} & \numprint{13.1} & \numprint{6.0} & \numprint{3.4} & \numprint{2.2}\\
benin & \numprint{188.1} & \numprint{47.9} & \numprint{12.2} & \numprint{5.5} & \numprint{3.2} & \numprint{2.1}\\
georgia & \numprint{322.1} & \numprint{83.6} & \numprint{22.5} & \numprint{9.8} & \numprint{5.6} & \numprint{3.6}\\
latvia & \numprint{355.2} & \numprint{92.0} & \numprint{23.3} & \numprint{10.6} & \numprint{5.9} & \numprint{4.0}\\
somalia & \numprint{420.1} & \numprint{108.7} & \numprint{27.6} & \numprint{12.6} & \numprint{7.1} & \numprint{4.6}\\
ethiopia & \numprint{825.9} & \numprint{215.7} & \numprint{53.9} & \numprint{24.4} & \numprint{13.9} & \numprint{9.0}\\
tunisia & \numprint{1200.1} & \numprint{303.1} & \numprint{77.7} & \numprint{34.6} & \numprint{19.7} & \numprint{12.9}\\
\bottomrule
\end{tabular}

\caption{Running time in seconds of \algo on the networks in
\Cref{tab:time-kt}.}
\end{table}
}{}

\end{document}